\documentclass[pdflatex,sn-nature]{sn-jnl}

\usepackage{graphicx}%
\usepackage{multirow}%
\usepackage{amsmath,amssymb,amsfonts}%
\usepackage{amsthm}%
\usepackage{mathrsfs}%
\usepackage[title]{appendix}%
\usepackage{xcolor}%
\usepackage{textcomp}%
\usepackage{manyfoot}%
\usepackage{booktabs}%
\usepackage{algorithm}%
\usepackage{algorithmicx}%
\usepackage{algpseudocode}%
\usepackage{listings}%
\usepackage{mhchem}
\usepackage{hyperref}
\usepackage{csquotes}
\usepackage{siunitx}
\usepackage{comment}
\hypersetup{
    colorlinks=true,
    linkcolor=blue,
    urlcolor=cyan,
    }

\theoremstyle{thmstyleone}%
\theoremstyle{thmstyletwo}%

\theoremstyle{thmstylethree}%

\begin{document}

\title[Title]{Bubble jetting in acoustic microdroplet vaporization}


\author*[1]{\fnm{Anunay} \sur{Prasanna}}\email{aanunay@ethz.ch}

\author[1]{\fnm{Samuele} \sur{Fiorini}}\email{sfiorini@ethz.ch}

\author[1,2]{\fnm{Gazendra} \sur{Shakya}}\email{gazendra.shakya@bracco.com}

\author*[1]{\fnm{Outi} \sur{Supponen}}\email{outis@ethz.ch}

\affil*[1]{\orgdiv{Institute of Fluid Dynamics}, \orgname{D-MAVT}, \orgaddress{\street{Sonneggstrasse}, \city{Z\"urich}, \postcode{8092}, \country{Switzerland}}}

\affil[2]{\orgname{Bracco Suisse SA}, \orgaddress{\street{Rte. de la Galaise 31, Plan-les-Ouates}, \city{Geneva}, \postcode{1228}, \country{Switzerland}}}

\abstract{Acoustic droplet vaporization denotes the phase-change of micron- and sub-micron-sized droplets upon the application of high-amplitude ultrasound. 
The asymmetric collapse of the incepted vapor bubbles within the droplets can give rise to high-speed liquid microjets.
Here, we describe acoustically-driven and bubble-pair jetting arising within the vaporizing droplet, observed experimentally with ultra-high-speed imaging at the microscale.
The existence of complex pressure fields due to the continued acoustic wave-droplet interaction and the nucleation of multiple bubbles within the droplet leads to rich dynamics, with the jets presenting behavioral self-similarity to millimetric bubbles under comparable conditions.
Evaporative instabilities that develop during bubble growth impede jet formation during bubble collapse.
Furthermore, the ability of the jets to pierce the droplet interface and penetrate into the surrounding fluid is discussed.
These powerful microjets could be harnessed to induce cell permeabilization for targeted drug delivery and treatment of cancerous tissue.}




\maketitle

\section{Introduction}
\label{sec:Intro}

Micron- and sub-micron-sized droplets are a class of acoustically-responsive agents that were first introduced in the 1990's to enhance contrast in ultrasound imaging~\citep{Sehgal1995SonographicMedia.,Albrecht1996RenalEchoGen,Forsberg1995ParenchymalAgent.}.
At body temperature, the droplets are typically in a superheated state and behave as microbubble precursors, since upon application of high-frequency ultrasound they undergo a phase-change process, commonly known as acoustic droplet vaporization (ADV)~\citep{kripfgans2000}.
The process shows great potential in therapeutic applications such as gas embolotherapy~\citep{Harmon2019MinimallyCarcinoma}, targeted drug delivery~\citep{chen2013,sheng2021}, and ablation techniques such as thermal ablation~\citep{Zhang2011AcousticUltrasound} and histotripsy~\citep{vlai2015}.

Sub-micron droplets provide greater circulation persistence than microbubbles in the human body~\citep{borden2020, Shakya2024Ultrasound-responsiveDelivery}, and their small size also allows them to extravasate into tumor tissue due to the enhanced permeability and retention (EPR) effect~\citep{Maeda2015TowardHeterogeneity}.
It has been hypothesized that the explosive bubble-droplet growth during ADV [$t \sim \mathrm{O}(10~\si{\micro\second})$], and the post-vaporization ultrasound-driven bubble dynamics [$t \sim \mathrm{O}(10 - 100~\si{\micro\second})$] can induce cell membrane permeabilization and death~\cite{fan2018}. 
However, the physical mechanisms associated with the initial instants of ADV [$t \leq \mathrm{O}(1~\si{\micro\second})$] that could also contribute to sonoporation remain unclear.
Therefore, a deeper understanding of the initial ADV physics is necessary to enhance control in medical applications, and to further optimize the process and make it biologically safe~\citep{Shakya2024Ultrasound-responsiveDelivery}.   

Considerable effort has been dedicated to interpret the physical mechanisms that can satisfactorily describe the initial ADV dynamics. 
Upon interacting with the ultrasound excitation, the spherical droplet behaves as an acoustic lens and amplifies the incoming wave, creating a pattern of standing waves that leads to complex pressure fields, with highly localized regions of negative pressure enabling vapor bubble nucleation~\cite{shpak2014,lajoinie2021,Fiorini2025PositiveVaporization}. 
High-speed visualization~\citep{shpak2013c, Fiorini2025PositiveVaporization, Abeid2024Ultra-high-speedPoint}, along with theoretical and numerical models~\citep{shpak2013, Ghasemi2022AnShell, Doinikov2014VaporizationValidation}, have been employed to describe the ensuing vapor bubble growth.
These models are based on the assumption of spherical symmetry, which strictly corresponds to vapor bubble nucleation occurring at the droplet center, and therefore, they are incapable of fully describing asymmetric bubble effects, which can initiate high-speed liquid jets that currently remain under investigated. 

Jetting in collapsing bubbles arises from a pressure gradient introduced through some form of asymmetry such as shock-waves~\citep{Sankin2005ShockBubbles} or ultrasound driving~\citep{Brujan2017JetsBoundary}, the presence of a surface (rigid~\citep{blake1981,Brujan2017JetsBoundary}, curved~\citep{tomita2002}, free~\citep{Blake1981GrowthSurface, Supponen2016ScalingBubbles}), or due to the influence of other bubbles~\citep{han2015,Mishra2022Flow-focusingBubbles}, and can essentially be described by the Kelvin impulse~\citep{Blake2015CavitationApplications}.
Acoustically-driven bubble jets have been especially studied within the context of employing bubbles for medicine. These jets are typically driven in the direction against the pressure gradient (maximum to minimum pressure) and depend on ultrasound parameters such as the driving amplitude, maximum bubble radius reached, and phase of the acoustic wave with respect to bubble inception~\citep{Brujan2005JetUltrasound, gerold2012, rossello2018}.

Recently, bubble jetting within millimetric droplets has been studied extensively due to the implications of explosions in confined liquid volumes for several industrial processes such as lithography~\cite{Banine2011PhysicalMicrolithography}, optical drop atomization~\cite{Lee2022AFiber}, and laser-based atmospheric monitoring techniques~\cite{Mei2015AtmosphericSystem}.
In particular, bubble nucleation within a droplet leads to coupled oscillations of the system, with the eccentricity of bubble location amplifying several types of jetting mechanisms, both in the nucleated bubbles and on the droplet surface~\citep{obreschkow2006, Rossello2023BubbleDroplet, Li2024CavitationFluid}.
Jets are also produced from the interactions of multiple bubbles, typically studied with laser-induced bubble pairs~\citep{sankin2010,han2015,Tomita2017PulsedTimes}. 
Their jetting properties, especially their direction and intensity, are controlled by the phase difference between the nucleation of the respective bubbles, their maximum size, and the relative separation distance between the two bubbles~\citep{han2015, Tomita2017PulsedTimes}.
Such jets have been proposed as a tool for creating highly flow-focused energy and material concentrators~\citep{sankin2010,Mishra2022Flow-focusingBubbles}.

Interestingly, ADV is a process that can present all the aforementioned driving mechanisms for jet formation.
It is activated by ultrasound and involves confinement effects due to the formation of vapor bubbles inside a droplet. 
Moreover, if a suitable ultrasound excitation is employed, multiple bubbles may appear within the droplet.
All of these factors imply strong and rapidly evolving pressure gradients, which could easily cause the nucleated vapor bubbles to undergo nonspherical collapses and rebounds.
Thermal effects can further complicate these interactions, since the droplets are typically in a superheated state.
As a mechanism to induce therapeutic effects in ADV, bubble jetting could be beneficial as it presents spatially localized, focused energy and fluid transport, similar to other cavitation bubble jets~\citep{sankin2010, Ohl2006SonoporationBubbles}.
If leveraged correctly, the piercing effects of these jets could improve drug and damage targeting, especially in ablation techniques. 
Therefore, we investigate the formation of ADV jets at the micrometric scale with ultra-high-speed imaging and elucidate their underlying mechanisms.

\section{Results}
\label{sec:Results}

\subsection{General dynamics}

\begin{figure}[t]
\centering
\includegraphics[width=\linewidth]{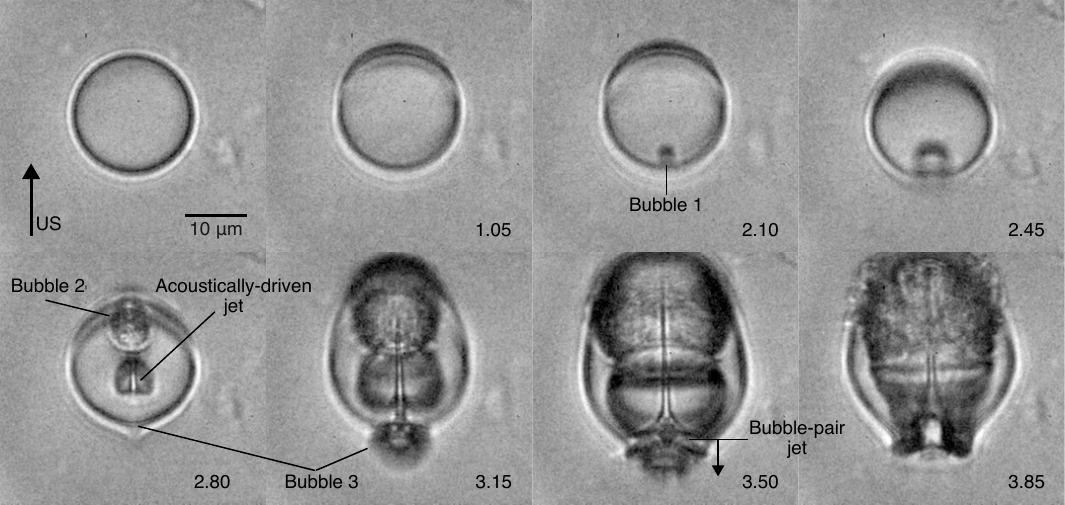}
\caption{\label{fig_1} \textbf{Snapshots from the ultra-high-speed recording of the ADV process.} The dynamics of vapor bubbles within a perfluoropentane (PFP, \ce{C5F12}) droplet of initial radius $R_\mathrm{d}=9.4$~\si{\micro\meter} ($T=37~\si{\celsius}$, $f_\mathrm{in}=3.5$~\si{\mega\hertz}) is depicted. The snapshots are labeled with their non-dimensional time, $t/t_\mathrm{d}$ ($t_\mathrm{d} = 0.285~\si{\micro\second}$). The ultrasound propagates from the bottom to the top of the image. The corresponding video is available as Supplementary Movie 1.}
\end{figure}

An example ADV process for a micrometric droplet visualized using ultra-high-speed imaging is depicted in Fig.~\ref{fig_1}. 
Here, the droplet has an initial radius of $R_\mathrm{d} = 9.4~\si{\micro\meter}$, is maintained at $T = 37~\si{\celsius}$ and sonicated at a frequency of $f_\mathrm{in}=3.5~\si{\mega\hertz}$, corresponding to a period of $t_\mathrm{in} = 0.285~\si{\micro\second}$. 
The first instant at which the incoming ultrasound interacts with the droplet is defined as $t=0$.

At $t/t_\mathrm{in} = 2.10$, a vapor bubble (labeled \textquote{Bubble 1} in Fig.~\ref{fig_1}) is nucleated inside the droplet, oscillates with the driving ultrasound and maintains its spherical shape during its first growth phase.
As the bubble collapses and rebounds, it translates away from the droplet interface and presents a needle-shaped jet (herein referred to as acoustically-driven jets), as seen at $t/t_\mathrm{in} = 2.80$.  
These are commonly the first type of microjets visualized in an ADV experiment and present velocities reaching up to 100~\si{\meter\per\second}, but lasting for a very short duration, typically $\tau_\mathrm{jet} \sim \mathrm{O}(0.2~\si{\micro\second})$.
As the bubble rebounds, the entrained jet continues to propagate and pierces the bubble's distal side, sometimes reaching the droplet interface and traveling into the surrounding fluid.
Acoustically-driven jets are further discussed in Sec.~\ref{sec:acWaveInDrop}.

Since ultrasound excitation is still active at this stage, several other vapor bubbles are nucleated at different times (labeled \textquote{Bubble 2} and \textquote{Bubble 3} in Fig.~\ref{fig_1} at $t/t_\mathrm{in} = 2.80,\;3.15$). 
These bubbles interact with each other and form jets (herein referred to as bubble-pair jets), with their dynamics further affected here by the proximity to the curved interface of the droplet.
\textquote{Bubble 3} (see Fig.~\ref{fig_1}) incepts what appears to be a jet, which merges with the existent, expanding acoustically-driven bubble jet.
Bubble-pair jets are further discussed in Sec.~\ref{sec:BP}.

\subsection{Acoustically-driven jets}
\label{sec:acWaveInDrop}

To evaluate how the transmitted wave and the resulting pressure gradients inside the droplet can drive vapor bubble jetting, we use a theoretical model described elsewhere~\citep{shpak2014,Fiorini2025PositiveVaporization}.
The main equations are given here for completeness, with further information available in Supplementary Note 1.
The acoustic pressure field inside a PFP droplet of density, $\rho_\mathrm{d}$, and speed of sound, $c_\mathrm{d}$, can be expressed as:
\begin{equation}
    \label{eq:pressure_transmission}
    p_\mathrm{t}(r,\theta,t) = \mathrm{Re} \left[\sum_{n=0}^\infty\sum_{m=0}^\infty a_{n} e^{i(n\omega_\mathrm{1} t + \phi_{n})} \alpha_{m,n} j_{m}(nk_2r) P_{m}(\cos\theta)\right] ,
\end{equation}
where the first summation represents the $n$ harmonics of the nonlinearly propagating source, $j_{m}$ is the spherical Bessel function of the first kind, $k_2 = \omega_1/c_\mathrm{d}$ is the fundamental wavenumber inside the droplet, and $P_m(\cos\theta)$ is the Legendre polynomial of order $m$.
$\alpha_{m,n}$ is defined as an amplification factor:
\begin{equation}
    \label{eq:alpha_mn}
    \alpha_{m,n} = (-i)^m (2m+1) \frac{j_{m}(k_1 R_\mathrm{d})h_{m}^{'(2)}(k_1 R_\mathrm{d}) - h_{m}^{(2)}(k_1 R_\mathrm{d})j_{m}^{'} (k_1 R_\mathrm{d})}{j_{m}(k_2 R_\mathrm{d})h_{m}^{'(2)}(k_1 R_\mathrm{d}) - \frac{\rho_\mathrm{w} c_\mathrm{w}}{\rho_\mathrm{d} c_\mathrm{d}} h_{m}^{(2)}(k_1 R_\mathrm{d})j_{m}^{'} (k_2 R_\mathrm{d})} ,
\end{equation}
where $\rho_\mathrm{w}$ is the density of water, $c_\mathrm{w}$ is the sound speed in water, $k_1 = \omega_1/c_\mathrm{w}$ is the fundamental wavenumber in the surrounding water, $h_{m} = j_{m} - iy_{m}$ is the spherical Hankel function of the second kind with $y_{m}$ being the spherical Bessel function of the second kind, and $'$ denotes differentiation with respect to $k_jR_\mathrm{d}$ with $j=1,2$. 

\begin{figure*}[!t]
\centering
\includegraphics[width=\linewidth]{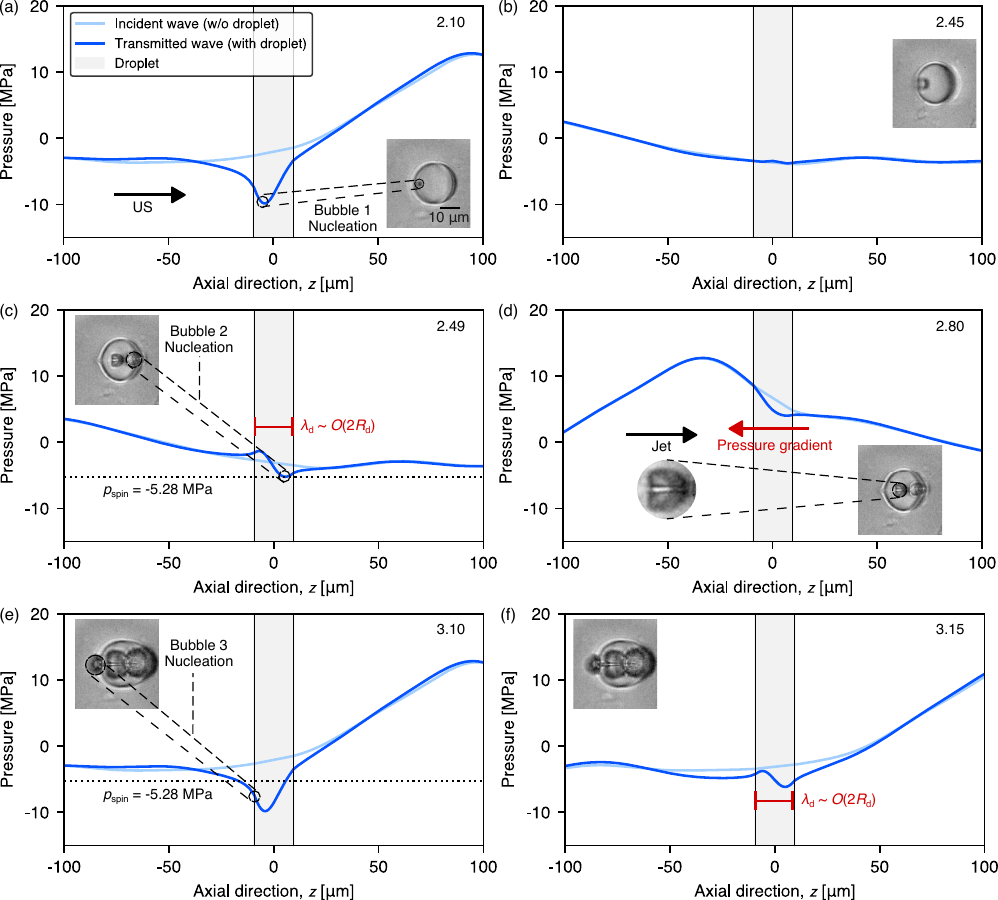}
\caption{\textbf{Computed pressure field along the droplet centerline for the undisturbed incident wave and the transmitted wave in the presence of the droplet, corresponding to the experiment depicted in Fig.~\ref{fig_1}}. The snapshots are labeled by their non-dimensional times, $t/t_\mathrm{in}$, and contain insets of the respective experimental frames. The ultrasound here propagates from left to right. The dotted line indicates the spinodal pressure of liquid PFP evaluated using the Redlich-Kwong equation of state~\cite{Qin2021PredictingVaporization}. 
(a) Instant of minimum pressure coinciding with the nucleation of \textquote{Bubble 1}. (b) Expansion of \textquote{Bubble 1}. (c) Instant of minimum pressure at the nucleation location of \textquote{Bubble 2}. The wave profile inside the droplet presents a wavelength, $\lambda_\mathrm{d} \sim \mathcal{O}(2R_\mathrm{d})$. (d) An asymmetric, steep pressure gradient within the droplet leads to the formation of a microjet. (e) Instant of minimum pressure at the nucleation location of \textquote{Bubble 3}. (f) Sustained existence of the acoustic jet created in (c), due to the presence of acoustic wavelengths of the order of the droplet radius. The corresponding animation for the steady-state acoustic wave solution is available in the Supplementary Movie 2.
\label{fig_2} }
\end{figure*}

The computed pressure field along the droplet centerline and the corresponding snapshots of the experiment in Fig.~\ref{fig_1} are depicted in Fig.~\ref{fig_2} for different time instants.
As the incoming ultrasound interacts with the spherical droplet, the acoustic impedance mismatch of the droplet (PFP) with its surroundings (water) causes the droplet to behave as an acoustic lens, focusing and amplifying the wave.
Fig.~\ref{fig_2}(a) shows the instant of minimum pressure that corresponds to the location of the first vapor bubble nucleus (labeled \textquote{Bubble 1} in Fig.~\ref{fig_1}). 
The pressure within the droplet is well below the spinodal limit of liquid PFP (calculated as $p_\mathrm{spin}(T = 37~\si{\celsius}) = -5.28~\si{\mega\pascal}$ from the Redlich-Kwong equation of state~\cite{Qin2021PredictingVaporization}) for sufficient time to trigger vapor nucleation.

Subsequently, the rarefaction phase of the incoming ultrasound enhances the negative pressure that consequently expands the vapor bubble as depicted in Fig.~\ref{fig_2}(b).
The focusing of the incoming multi-frequency acoustic wave, coupled with the multiple internal reflections, lead to standing wave profiles within the droplet with a typical wavelength of the order of the droplet diameter, $\lambda_\mathrm{d} \sim \mathrm{O}(2R_\mathrm{d})$, which is noticeably shorter than the wavelength of the incoming ultrasound.
Occasionally, these wave profiles present negative pressures below the spinodal limit of liquid PFP, therefore allowing for the nucleation of new vapor bubbles, such as \textquote{Bubble 2} in Fig.~\ref{fig_2}(c). 
The combination of these oscillating standing wave profiles (see Supplementary Movie 2), followed by the incoming compression phase of the ultrasound, creates a steep spatial pressure gradient within the droplet, leading to the collapse of the vapor bubble that presents a well-developed liquid microjet upon rebound (see Fig.~\ref{fig_2}(d)).

Fig.~\ref{fig_2}(e) presents the same profile as Fig.~\ref{fig_2}(a), since it coincides with the completion of one acoustic cycle, enabling the nucleation of \textquote{Bubble 3}, and the next instant in Fig.~\ref{fig_2}(f) again shows a standing wave pattern within the droplet that enhances the continued growth of the acoustically-driven jet.
These results indicate that the complex, rapidly varying acoustic fields within the droplet provide spatial pressure gradients over distances comparable to the maximum bubble radius that are steep enough to cause the bubble to jet, sometimes even before a full ultrasound cycle has passed.
This observation corresponds well with other studies involving ultrasound-driven millimetric bubbles that have reported jetting when the maximum bubble radius is approximately equivalent to the incoming acoustic wavelength, which implies that a noticeable spatial gradient across the bubble surface is required to create jets~\citep{gerold2012,rossello2018}.
Furthermore, the bubble dynamics could induce additional acoustic contributions not considered here that may further enhance the formation of jets and vapor bubble nucleation.



To provide better insight on the vapor bubble collapse and jetting, we model the radial dynamics of a spherically symmetric bubble located in an infinite medium until its first collapse using a modified Rayleigh-Plesset-type equation~\citep{hao1999,stricker2011,shpak2013}:
\begin{equation}
    \left(1 - \frac{\dot{R}_\mathrm{b}}{c_\mathrm{d}}\right)R_\mathrm{b}\ddot{R}_\mathrm{b} + \frac{3}{2}\left(1 - \frac{\dot{R}_\mathrm{b}}{3c_\mathrm{d}}\right)\dot{R}_\mathrm{b}^2 = \frac{1}{\rho_\mathrm{d}}\left(1 + \frac{\dot{R}_\mathrm{b}}{c_\mathrm{d}}+\frac{R_\mathrm{b}}{c_\mathrm{d}}\frac{d}{dt}\right)[p_\mathrm{b}(t) - p_\mathrm{a}(t)],
    \label{eq:KM}
\end{equation}
where $R_\mathrm{b}$ is the bubble radius, and dots denote differentiation with respect to time.
The bubble pressure is denoted by $p_\mathrm{b}$ and is evaluated by coupling the Clausius-Clapeyron equation with an advection-diffusion equation for the evolution of the bubble surface temperature. 
Further details on the modeling are available in Supplementary Note 2. 
\begin{figure}[!t]
    \centering
    \includegraphics[width=0.6\linewidth]{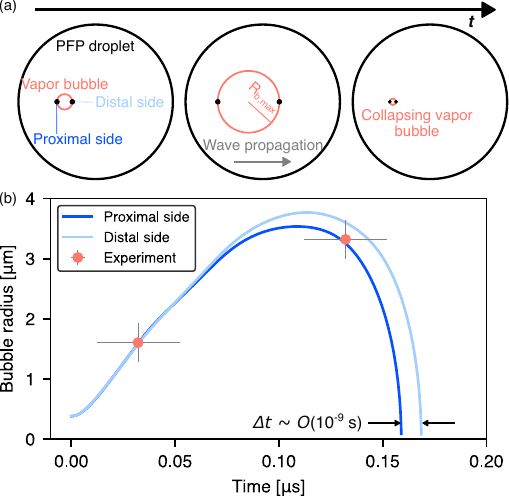}
    \caption{\textbf{Radial dynamics of the proximal and distal side of the vapor bubble corresponding to \textquote{Bubble 1} in Fig.~\ref{fig_1}}. (a) Schematic depicting the bubble dynamics modeled over time, with labels indicating the proximal and distal side of the bubble wall with respect to the ultrasound. (b) Bubble radius over time computed using Eq.~\eqref{eq:KM} for the proximal and distal side compared with the experiment for the bubble labeled \textquote{Bubble 1} in Fig.~\ref{fig_1}. Typically, the proximal side collapses faster than the distal side, with a time difference of the order of $\Delta t \sim O(10^{-9} \;\mathrm{s})$, which corresponds to an asymmetric collapse and the formation of a jet. The error bars represent the measurement error in bubble radius due to image processing and the error in the recording time due to the camera frame rate used.}
    \label{fig_3}
\end{figure}

The pressure, $p_\mathrm{a} = p_\infty + p_\mathrm{acoustic}(t)$, refers to the external driving pressure on the bubble wall.
The acoustic pressure, $p_\mathrm{acoustic}(t)$, is evaluated here at two different locations on the bubble wall [see Fig.~\ref{fig_3}(a)], corresponding to the proximal and distal side of the bubble with respect to the incoming ultrasound along the droplet centerline.
The values of $p_\mathrm{acoustic}(t)$ and their time derivatives at the two locations are obtained by interpolating the solution of the transmitted acoustic pressure from Eq.~\eqref{eq:pressure_transmission}, evaluated on a coarser grid.
The initial bubble radius is set by assuming that the incepted nucleus has to work initially against the surface tension to grow, $R_\mathrm{b,0} = 2\gamma/[p_\mathrm{v}(T_\mathrm{\infty}) - p_\mathrm{v}(T_\mathrm{b})]$, as described in Plesset and Prosperetti~\cite{Prosperetti1978Vapour-bubbleLiquid}.
With this assumption, the initial bubble velocity is set to zero.
The solution of Eq.~\eqref{eq:KM} for the experiment in Fig.~\ref{fig_1} is plotted in Fig.~\ref{fig_3}(b).

The starting phase of the acoustic wave within the droplet is varied to match the bubble radii measured experimentally from the high-speed videos (see Supplementary Figure S2).
The adjustment is necessary due to the stochastic nature of nucleation and the insufficient temporal resolution to measure the exact phase of nucleation with respect to an ultrasound cycle. 
From its inception, the bubble is driven for one cycle and typically collapses within that cycle.
Fig.~\ref{fig_3}(b) shows that the proximal side of the bubble wall collapses faster than the distal side, with the bubble being subjected to steepening pressure gradients over time during its collapse, similar to the profiles depicted in Fig.~\ref{fig_2}(d).

The short but noticeable time difference between the collapse times of the two points, typically of the order of $\Delta t \sim \mathrm{O}(10~\si{\nano\second})$, can be attributed to the acoustic \textquote{lag} in the pressure acting on the two points.
This effect is enhanced in PFP compared to water due to its lower speed of sound ($c_\mathrm{d} = 406~\si{\meter\per\second}$ as compared to $c_\mathrm{w} = 1481~\si{\meter\per\second}$), which implies that for the same pressure driving, the spatial gradient is approximately $c_\mathrm{w}/c_\mathrm{d} \approx 3.65$ times larger in PFP than in water. 
Additionally, the pressure is further amplified by the lensing behavior of the droplet, steepening the gradient within the droplet. 
Therefore, it can be concluded that the spatial pressure gradients within the droplet induce the \textquote{time lag}($\Delta t$) corresponding to a jet directed from the faster-collapsing side towards the slower side.

The features of the acoustically-driven jet depicted here are similar to the one described by Longuet-Higgins and Oguz~\cite{longuet1995} for high-speed, inward jets typically noticed during asymmetric bubble collapse in the presence of external pressure gradients, and has also been seen in other acoustically-driven bubble jets~\citep{gerold2012, rossello2018}.
These jets can transport encapsulated fluid mass as they penetrate into the host fluid and exert dynamic pressures on interfaces, $p_\mathrm{dyn} \approx \frac{1}{2}\rho_\mathrm{d}u_\mathrm{avg}^2$ (where $\rho_\mathrm{d}$ is the density of PFP, and $u_\mathrm{avg}$ is the average jet velocity~\citep{Brujan2017JetsBoundary}), which for the case depicted in Fig.~\ref{fig_1} corresponds to $p_\mathrm{dyn} \approx 7.3~\si{\mega\pascal}$.


\subsection{Bubble-pair jets}
\label{sec:BP}

\begin{figure*}[!t]
\centering
\includegraphics[width=\linewidth]{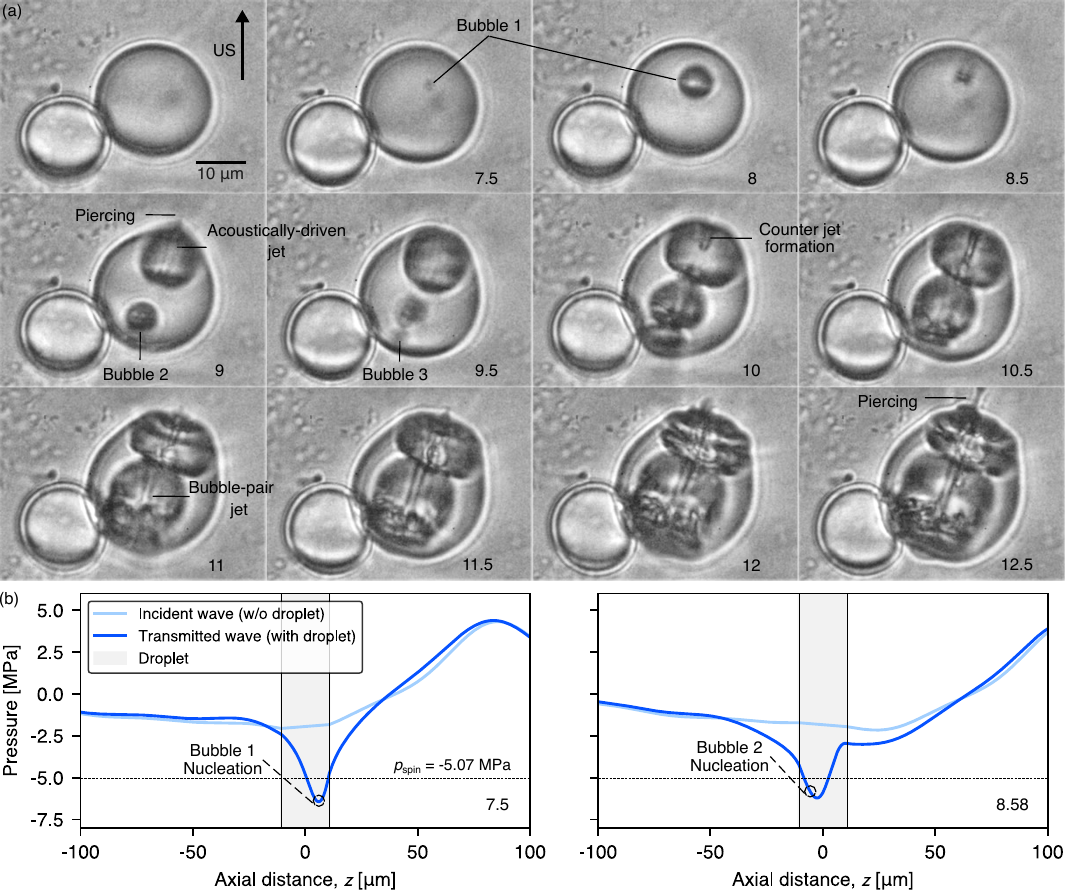}
\caption{\label{fig_4} \textbf{Nucleation of multiple bubbles and the formation of bubble-pair jets in a vaporizing PFP droplet.} (a) Snapshots from an ultra-high-speed recording of an ADV process and subsequent collapse dynamics of incepted vapor bubbles. The initial radius of the PFP droplet is $R_\mathrm{d}=10.7$~\si{\micro\meter} ($T=39~\si{\celsius}$, $f_\mathrm{in}=5$~\si{\mega\hertz}). The snapshots are labeled with their non-dimensional time, $t/t_\mathrm{in}$. The corresponding video is available as Supplementary Movie 3. (b) Steady-state solution of the traveling acoustic wave inside the droplet shown in (a), obtained for different instants by solving Eq.~\eqref{eq:pressure_transmission}. The snapshots are labeled by their non-dimensional time, $t/t_\mathrm{in}$, and show instants when each vapor bubble is expected to nucleate in (a). The instant for Bubble 3 is not provided due to the uncertainty in nucleation position and the acoustic field within the droplet. The spinodal pressure evaluated from the Redlich-Kwong equation~\citep{Qin2021PredictingVaporization} is depicted for reference. The animation for the steady-state acoustic wave solution is available as Supplementary Movie 4.}
\end{figure*}

Multiple vapor bubbles can be nucleated within a droplet during ADV, and their interaction can lead to bubble-pair jets.
An example is depicted in Fig.~\ref{fig_4} for a droplet of size, $R_\mathrm{d}  = 10.7~\si{\micro\meter}$, driven by a 5~\si{\mega\hertz} wave. 
The incoming wave is focused by the droplet and incepts a vapor bubble on the distal side, as seen in Fig.~\ref{fig_4}(a).
The steady-state acoustic solution in Fig.~\ref{fig_4}(b) at $t/t_\mathrm{in} = 7.5$ shows a pressure minimum well below the spinodal limit of PFP ($p_\mathrm{spin}(T = 39~\si{\celsius}) = -5.07~\si{\mega\pascal}$) at the location of nucleation.
\textquote{Bubble 1} grows, and on its rebound, presents an acoustic jet as described in Sec.~\ref{sec:acWaveInDrop}. 
At $t/t_\mathrm{in} = 9$, the expansion of the rebounding bubble along with the entrained jet pierces the droplet and is followed by the reversal of the jet in the next cycle.
This counter-jet can arise due to the impact of the original jet on the distal side of the bubble wall~\cite{Supponen2016ScalingBubbles} and is further enhanced here by the proximity of the droplet interface to the growing bubble and the rapid change in direction of the acoustic pressure gradient within the droplet.
In ADV, this phenomenon usually occurs when the initial acoustically-jetting bubble is nucleated on the distal focus of the droplet, as depicted in Fig.~\ref{fig_4}.
These counter jets are typically slower with a slightly longer lifetime [$u_\mathrm{avg} \approx 30~\si{\meter\per\second}$, $\tau_\mathrm{jet} \sim \mathrm{O}(0.5~\si{\micro\second})$] and are directed towards the droplet center.
It must be noted that unlike other studies involving cavitation in millimetric droplets~\citep{Rossello2023BubbleDroplet, Reese2024CavitationSurfaces, Li2024CavitationFluid}, the droplet interface is not a key driver for jets in our experiments (see Supplementary Note 3 for further details).


At $t/t_\mathrm{in} = 9$, a second bubble is nucleated at the proximal focus inside the drop [labeled \textquote{Bubble 2} in Fig.~\ref{fig_4}(a)].
As this bubble grows, a third bubble (labeled \textquote{Bubble 3}), slightly out-of-phase with \textquote{Bubble 2}, is nucleated nearby.
It is predicted that both bubbles nucleate within a few nanoseconds of each other, although the experimental spatiotemporal resolution is insufficient for \textquote{Bubble 3} to be visible until $t/t_\mathrm{in} = 9.5$.
The initiation of bubble-pair jets is marginally visible in the snapshots at $t/t_\mathrm{in} = 10, 10.5$.
The bubble-pair jets arise as a consequence of flow-focusing between the two bubbles: the differing expansion rates of the out-of-phase bubbles forces the fluid between them into one of the bubbles~\citep{han2015,Mishra2022Flow-focusingBubbles}.
In Fig.~\ref{fig_4}(a), this corresponds to \textquote{Bubble 3} pushing the liquid above into \textquote{Bubble 2}.
At later stages in Fig.~\ref{fig_4}, the bubble-pair jet merges with the existing jets within the droplet and pierces the droplet interface, penetrating well into the host fluid.
This bubble-pair jet has an average velocity of $u_\mathrm{avg} \approx 78~\si{\meter\per\second}$ and applies an average dynamic pressure of $p_\mathrm{dyn} \approx 4.7~\si{\mega\pascal}$ upon impacting the droplet interface.

\begin{figure}[!t]
    \centering
    \includegraphics[width=0.5\linewidth]{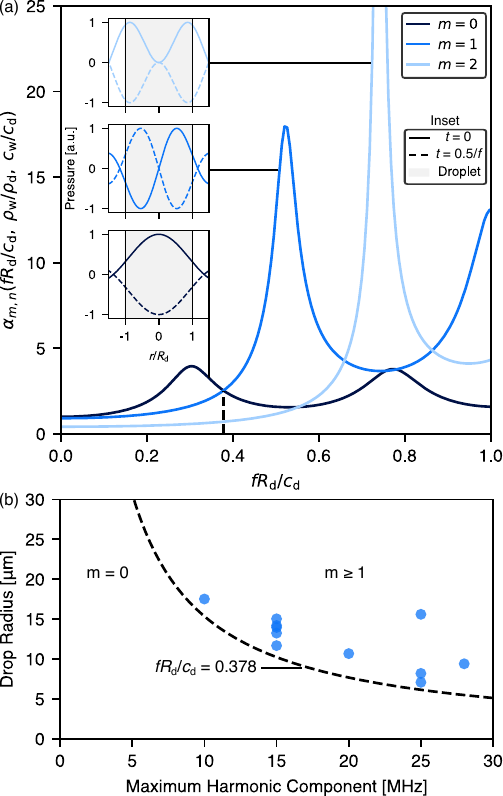}
    \caption{\label{fig_5} \textbf{Variation of amplification factors as a function of the wavenumber inside the droplet and dependence of bubble-pair jetting on higher harmonics.} (a) The variation of the first three modes of the amplification factor, $\alpha_{m,n}$, given by Eq.~\eqref{eq:alpha_mn}, as a function of the generalized wavenumber for a PFP droplet in water. The dashed line indicates the value of $fR_\mathrm{d}/c_\mathrm{d}$ beyond which the higher modes present larger amplification than the mode, $m=0$. The insets show the typically expected extrema pressure profiles for a given dominant mode. The corresponding animations for the different modes are available as Supplementary Movies 5 -- 7. (b) The droplet radius and the maximum significant harmonic component of the incoming wave are plotted for the droplets that present bubble-pair jetting. The dashed line represents the value above which the higher modes are dominant. Note that only the highest harmonic is shown for the viewer's convenience, although all the lower harmonics are also excited for each case. The measured drop radii have an error of $\pm 0.32~\si{\micro\meter}$, and the maximum measured harmonic is limited by the bandwidth of the hydrophone ($f_\mathrm{max} = 30~\si{\mega\hertz}$).}
\end{figure}

\begin{figure}
    \centering
    \includegraphics[width=0.5\linewidth]{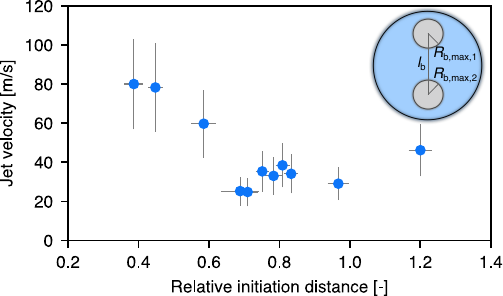}
    \caption{\label{fig_6} \textbf{The average jet velocity, $u_\mathrm{avg}$, shown as a function of the relative initiation distance, $l^*$, for ADV bubble-pair jets.} The inset schematic defines the dimensional variables needed to evaluate the relative initiation distance. Error bars are obtained by a Monte-Carlo approximation on the measured pixels for the bubble radii, bubble separation and jet tip distances, and the interframe time.}
\end{figure}

It is clear that having multiple regions with significant negative pressure within the droplet promotes the nucleation of closely spaced vapor bubbles that can induce bubble-pair jetting.
Locations of pressure minima within the droplet can be predicted by comparing the amplification factors for different modes, $m$, from Eq.~\eqref{eq:alpha_mn}, with respect to the dimensionless wavenumber, $fR_\mathrm{d}/c_\mathrm{d}$, inside the droplet as depicted in Fig.~\ref{fig_5}(a)~\citep{Fiorini2025PositiveVaporization}. 
Here, $f$ represents a generic frequency that incorporates the fundamental $ f_0 $ and the higher harmonics, $n f_0$, of the incoming wave.  
If the most amplified mode is $m=0$, then the global pressure extrema are expected at the droplet center, also called the \textquote{resonance} regime (see inset of Fig.~\ref{fig_5}(a))~\citep{lajoinie2021}. 
Similarly, significant amplification of modes $m \geq 1$ presents at least two focal points within the droplet, which corresponds to the \textquote{focusing} regime~\citep{shpak2014}. 
Representative pressure extrema within the droplets are depicted in the inset of Fig.~\ref{fig_5}(a), and the dashed line indicates the value of $fR_\mathrm{d}/c_\mathrm{d}$ above which the higher modes, $m\geq1$, likely dictate the wave dynamics inside the droplet. 

Higher modes are typically amplified by the higher harmonics of incoming acoustic waves, since they present larger wavenumbers within micrometric droplets~\cite{Fiorini2025PositiveVaporization}.
A routine Fast Fourier Transform (FFT) is performed on the experimentally measured pressure signals to determine their harmonics with a significant contribution [$a_n \geq 0.05a_0$ as per eq.~(\ref{eq:pressure_transmission})] (see Supplementary Note 4 for full details). 
Transferring the value of the generic wavenumber in Fig.~\ref{fig_5}(a) on to the $(f, R_\mathrm{d})$ plane in Fig~\ref{fig_5}(b), it can be seen that the highest harmonic components for the cases where bubble-pair jetting occur indeed lie above the value of $fR_\mathrm{d}/c_\mathrm{d} = 0.378$.
This implies that although all the corresponding lower harmonics are also excited for each case indicated in Fig.~\ref{fig_5}(b), having a noticeable harmonic contribution that can amplify an higher mode is important to nucleate multiple bubbles. 
While this is a necessary condition to ensure the creation of clearly distinguishable bubble-pair interactions, it is not a sufficient condition for bubble-pair jetting.

The shape, speed, and direction of the jets resulting from such bubble-bubble interactions are similar to what has been observed for millimetric bubble pairs~\citep{han2015,Tomita2017PulsedTimes}.
They also depend on similar parameters, including the relative sizes of bubbles, the phase difference of bubble nucleation, and the separation distance between them. 
A relevant parameter that can be evaluated to characterize the jet speed is the relative initiation distance~\citep{han2015}:
\begin{equation}
    l^* = \frac{l_\mathrm{b}}{R_\mathrm{b,max,1}+R_\mathrm{b,max,2}},
    \label{eq:l_b}
\end{equation}
where $l_\mathrm{b}$ is the separation distance between the two bubbles, and $R_{\mathrm{b,max,}j}$ is the measured maximum bubble radius prior to jet formation for the bubbles with $j = 1,2$.
Fig.~\ref{fig_6} plots the measured average jet velocity versus the relative initiation distance for the observed bubble-pair cases.
Similar to millimetric bubble pairs reported in the past, it can be noted that the flow-focusing effect is more enhanced if the bubbles are nucleated closer together. 
These jets have a longer lifetime than the other types of jets observed in ADV [typically $\tau_\mathrm{jet} \sim \mathrm{O}(1~\si{\micro\second})$]. 

In our ADV cases, the bubble pairs are normally out of phase with each other due to their inception from the continuously varying wave profiles within the droplet.
The steep pressure gradients combined with acoustic wavelengths comparable to the droplet size present an interesting example of acoustically generated out-of-phase vapor bubble-pair jetting in a small confined volume, which is a phenomenon not typically observed in acoustic bubble cavitation. 
The phase difference and proximity of these out-of-phase bubbles is essential in promoting thick and powerful bubble-pair jets directed outwards in ADV.
Such jets, usually called \textquote{catapult} jets~\cite{Chew2011InteractionField}, are capable of transporting large amounts of material and could potentially be useful in therapeutic applications.
In-phase bubble pairs probably occur in ADV when bubble clouds are nucleated instead of single detectable bubbles and probably play a minimal role in applications as they would be directed towards each other inside the droplet.
However, observing jetting in such cases is challenged by the imaging resolution.  

\subsection{Conditions for visible jetting} 



\begin{figure*}[!t]
    \centering
    \includegraphics[width=\linewidth]{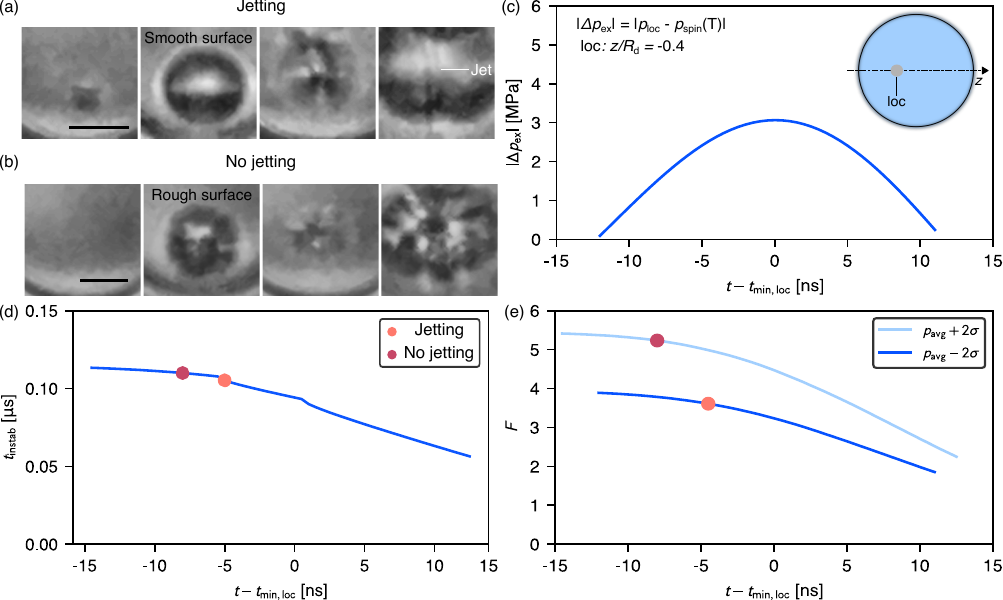}
    \caption{\textbf{The role of the evaporative instability in indicating the likelihood of bubble jetting. Two different vaporization events with the same physical parameters ($R_\mathrm{d} \approx 9.2~\si{\micro\meter}$, $f_\mathrm{in} = 5~\si{\mega\hertz}$, $P_\mathrm{pnp} \approx 4~\si{\mega\pascal}$, and $\Delta T_\mathrm{s} \approx 11~\si{\celsius}$) are compared.} (a) Zoomed-in snapshots of a vaporization event, presenting a bubble growing with a \textquote{smooth} surface, allowing it to form a microjet during collapse. The scale bar corresponds to 5~\si{\micro\meter} and the interframe time is 100~\si{\nano\second} ($t/t_\mathrm{in} = 0.5$). (b) Similar to (a), but for a non-jetting case, where the bubble exhibits a \textquote{rough} surface and is unstable during its growth. (c) The magnitude of the excess pressure below the spinodal limit of PFP at $T=40~\si{\celsius}$ in a given ultrasound cycle, at a fixed location and only for instants where this value is larger than zero. (d) The total time of instability experienced by a bubble that is nucleated at a given phase of the ultrasound. The expected instability times for the jetting and the non-jetting cases are marked. (e) The figure of merit, $F$, given by Eq.~\eqref{eq:FOM} for the two driving signals, $p_\mathrm{avg} \pm 2\sigma$. The time axis in the graphs above (c -- e) is given such that $t=0$ corresponds to the instant when the minimum pressure is achieved at the location of interest ($z/R_\mathrm{d} = -0.4$).}
    \label{fig_7}
\end{figure*}

Despite maintaining all physical parameters the same, there were noticeable differences in the observed vapor bubble dynamics and consequently in the occurrence of jets, with two examples depicted in Fig.~\ref{fig_7}.
Here, we classify a jet as a single high-speed liquid stream piercing a bubble during its collapse.
In the jetting case [Fig.~\ref{fig_7}(a)], the bubble surface is smooth during its growth phase, while the bubble in the non-jetting case [Fig.~\ref{fig_7}(b)] presents a highly perturbed surface.
We hypothesize that these perturbations arise due to the fact that the phase change of the nucleated bubble is still in progress~\citep{wong2011}.
Phase-change-driven instabilities of vapor bubbles within droplets have been observed in experiments performed at the superheat limit of liquids~\citep{Shepherd1982RapidLimit,Frost1986EffectsLimit}.
The instability is characterized by a \textquote{crinkling} or \textquote{roughening} of the bubble interface and has been reported in vapor explosions initiated by droplet heating with lasers~\citep{Xie1991EvaporativeDroplets}, and vaporization using sound~\citep{Apfel1975AcousticallyDroplets} and shock waves~\citep{Frost1989ExperimentsWave} in droplets at a larger scale [$R_\mathrm{d} \sim \mathrm{O}(1~\si{\milli\meter})$]. 

In ADV, the mass flux across the vapor-liquid interface is substantial due to large radial bubble velocities during its growth. 
This is especially true in the initial stages, when the incoming ultrasound is still active after bubble nucleation.
Changes in pressure have been shown to play a significant role in enhancing or suppressing the instability in evaporative processes of superheated liquids~\citep{Palmer1976ThePressure,Frost1986EffectsLimit}.
To estimate the stability of bubble growth, we employ the modified planar theory for evaporating interfaces in superheated liquids and obtain a dispersion relation~\citep{Frost1986EffectsLimit}:
\begin{equation}
    \Omega^2 + \frac{2\chi K}{\chi+1}\Omega + \frac{1}{\chi+1}\left[\chi(\chi-1)K^2 + \frac{2N_\mathrm{w}K + K^3}{2N_\mathrm{i}}\right] = 0,
    \label{eq:LD_Instability}
\end{equation}
where $\Omega = \omega R_\mathrm{b}/\dot{R}_\mathrm{b}$ is the dimensionless growth rate, $K = kR_\mathrm{b}$ is the dimensionless wavenumber, $\chi = \rho_\mathrm{v}/\rho_\mathrm{d}$ is the density ratio of the PFP vapor to its liquid, $N_\mathrm{w} = (\rho_\mathrm{d} - \rho_\mathrm{v})R_\mathrm{b}^2\ddot{R}_\mathrm{b}/2\gamma$ is the Weber number, and $N_\mathrm{i} = \rho_\mathrm{d}R_\mathrm{b}\dot{R}_\mathrm{b}^2/2\gamma$ is defined as the \textquote{inertia} number~\citep{Frost1986EffectsLimit}.
Fundamentally, $N_\mathrm{w}$ and $N_\mathrm{i}$ contrast the acquired inertia of the liquid near the bubble due to its acceleration and velocity with respect to the surface tension.
From Eq.~\eqref{eq:LD_Instability}, the condition for the bubble interface to become unstable can be derived as: 
\begin{equation}
    \dot{R}_\mathrm{b}^4 > \frac{4\gamma\ddot{R}_\mathrm{b}}{\chi^2 (\rho_\mathrm{d} - \rho_\mathrm{v})} \cdot
    \label{eq:unstab}
\end{equation}
At all times where the bubble growth is unstable in our experiment, the perturbation wavelengths satisfy the geometric limit of the bubble ($\lambda_\mathrm{u} < 2\pi R_\mathrm{b}$).
Effectively, the bubble is only unstable when the radial acceleration is negative, which occurs within a certain time frame after the initial stages of surface tension-dominated growth.
For the two cases presented in Fig.~\ref{fig_7}(a) and~\ref{fig_7}(b), the bubble dynamics is modeled by combining Eq.~\eqref{eq:pressure_transmission} and Eq.~\eqref{eq:KM}.
For a single ultrasound cycle, we select the phases at which nucleation can occur by evaluating the excess pressure below the spinodal limit at a given location within the droplet, $|\Delta p_\mathrm{ex}| = |p_\mathrm{loc} - p_\mathrm{spin}(T)| \ \forall \ p_\mathrm{loc} < p_\mathrm{spin}(T)$ (here, $p_\mathrm{spin}(T = 40~\si{\celsius}) = -4.99~\si{\mega\pascal}$), so that nucleation is expected only when $|\Delta p_\mathrm{ex}| > 0$. 
The variation of $|\Delta p_\mathrm{ex}|$ with the phase of the ultrasound at $z/R_\mathrm{d} = -0.4$ is given in Fig.~\ref{fig_7}(c).

For each instant that nucleation is possible, the dynamics of the growing bubble can be modeled as described in Sec.~\ref{sec:acWaveInDrop}, and the total duration of instability can be estimated from the condition provided in Eq.~\eqref{eq:unstab}.
Fig.~\ref{fig_7}(d) shows that the duration of instability decreases for later nucleation timings within the window of negative pressure in a given region, indicating that spending a shorter average duration experiencing expansion is beneficial for bubble stability. 
Considering the almost identical times of instability of the two cases shown in Fig.~\ref{fig_7}(a) and (b), it is unclear what causes the difference in the bubble dynamics.
We suspect that the differing dynamics could be attributed to burst-to-burst pressure variations of the applied signal (see Supplementary Note 5).
Within consecutive shots (here, $N=5$), the experimentally measured incident pressure shows relatively minor deviations [$\sigma \sim \mathrm{O}(100~\si{\kilo\pascal})$].
However, when these variations are amplified within the droplet, they lead to rather large pressure differences at a given location.
Provided there is a standard deviation ($\sigma$) on the amplitude of the average applied pressure, the pressure differences within the droplet for the two extreme excitations, $p_\mathrm{avg} \pm 2\sigma$, are on the order of \si{\mega\pascal} (see Supplementary Figure~S5).
If the driving pressure felt by the growing bubble is more negative (corresponding to an applied signal of amplitude $p_\mathrm{avg} + 2\sigma$), the growth rate of the instability is higher, increasing the probability of the bubble surface roughening.

To address the dependence of both the growth rate and the time of instability on bubble jetting, we introduce a modified metric, similar to the figure of merit used in previous studies~\citep{Shepherd1982RapidLimit,Frost1986EffectsLimit}.
\begin{equation}
    \label{eq:FOM}
    F = \int_{t_\mathrm{i,s}}^{t_\mathrm{i,f}} \omega' dt' ,
\end{equation}
where $\omega'$ indicates the maximum predicted growth rate at every instant of bubble growth, and $t_\mathrm{i,s}$ and $t_\mathrm{i,f}$ represent the start and end time of the instability.
This metric takes into account the fluctuating growth rates over time due to the rapid temporal pressure changes.
The figure of merit and its dependence on the phase of ultrasound driving is compared in Fig.~\ref{fig_7}(e) with the corresponding experimental points plotted for reference, where we assume that stable bubble growth leading to jetting is driven by the input signal $p_\mathrm{avg} - 2\sigma$, and unstable bubble growth is driven by the signal $p_\mathrm{avg} + 2\sigma$.
This choice is made as larger values of $F$ indicate higher probability of bubble instability.

Eq.~(\ref{eq:FOM}) could explain the differences in our experiments, with lower values of $F$ representing stable bubble growth, allowing the formation of a microjet upon rebound, whereas higher values of $F$ indicate the presence of higher order perturbations on the bubble surface, masking jet formation. 
The argument for the evaporative instability dictating jet formation is further strengthened by the fact that visible jets are only observed when the vaporization occurs near its threshold. 
The vaporization threshold is defined here as the lowest peak negative pressure (PNP) required to nucleate a vapor bubble within a given droplet~\cite{aliabouzar2018}.
Furthermore, this implies that visualizing jets and monitoring the metric $F$ that incorporates the effect of several physical parameters [$F = \Pi(p_\mathrm{pnp}, f,  R_\mathrm{d}, \phi, R_\mathrm{b,0}, \Delta T_\mathrm{s} )$] can help gain some insight into predicting initial nucleation dynamics, which is a topic that still remains under investigated.
However, the phenomenon is highly sensitive to minor changes in external conditions~\citep{Frost1986EffectsLimit,Frost1989ExperimentsWave,Xie1991EvaporativeDroplets}, which implies that the transient effects of the driving acoustic signal could also play a significant role on the instability onset.

While we investigate the effects of the nucleation phase with respect to a single steady-state cycle, it must be noted that nucleation can also be initiated by the transient acoustic cycles and does not occur consistently at the same cycle of the 10-cycle pulse (5--7 cycles on average for similar conditions as in Fig.~\ref{fig_7}).
Indeed, while we classify nucleation probability only using the variable $|\Delta p_\mathrm{ex}|$, it would be prudent to also take into account the effect of the total time a region spends below the spinodal limit, which would provide a more nuanced explanation of the possibility of jetting.
However, modeling nucleation in these rapidly changing pressure fields is a tedious task and is out of the scope of the present work.

\section{Discussion}
\label{sec:disc}

We performed a total of 160 vaporization events by varying the applied frequency, pressure, and the degree of superheat over a wide range of droplet sizes.
Out of these, about 51 droplets presented jetting of some form, with 15 droplets showing multiple instances of jetting.
Since piercing the droplet interface could allow for sonoporation, it is observed that acoustically-driven jets pierce the droplets approximately 28\% of the times and bubble-pair jets about 64\% of the time. 
Typically, acoustically-driven jets are the fastest jets in ADV, but are mostly contained within droplets and do not travel deep into the host fluid upon piercing. 
Bubble-pair jets are initiated with high-speed, are long-lived, and show the maximum penetration capability of all jet types.   

It is difficult to determine the exact conditions under which piercing can occur.
However, since bubble-pair jets show pronounced piercing with their extended lifetime, it is beneficial to induce acoustic focusing within the droplet to harness these jets.
Our findings indicate that by selecting suitable wavenumbers (i.e., higher acoustic frequencies, larger droplets), higher amplification modes could be excited within a droplet (see Fig.~\ref{fig_5}), providing statistically higher chances of multiple bubble nucleation and therefore, bubble-pair jetting.
Focusing could also be beneficial to promote piercing acoustic jets, since they mostly occur when bubbles are nucleated at the distal focus within the droplet core that subsequently jet towards the droplet interface.
The phase of bubble nucleation with respect to the incoming ultrasound could be a key parameter in determining whether an acoustic jet pierces the interface.
Acoustic jets are generally enhanced when the compression phase of the incoming acoustic wave coincides with the maximum size of the bubble~\citep{Sankin2005ShockBubbles,rossello2018}. 
However, correctly identifying the phase of nucleation in our experiments is extremely difficult.
Additionally, the continuous acoustic driving of these jets forces them to grow along with the bubble and further enhances their piercing capability.
While piercing jets can exert high pressure on their surroundings, they are short-lived and occur scarcely in ADV. 
The benefit of jets is that they offer precise targeting enabling for example, better delivery of encapsulated drugs.
Moreover, their spatially localized behavior can present a high degree of energy focusing that can be beneficial in ablation techniques.

Finally, it is worth discussing the implications of our results for sub-micrometric droplets that are more relevant in biomedical applications.
It is extremely difficult to verify whether we obtain jets and visualize their subsequent behavior for droplets of smaller sizes, due to experimental limitations such as optical diffraction.
If jets are present, the self-similarity of jetting bubbles is expected to hold as we reduce droplet size, which implies that the piercing capabilities of these jets could still be utilized.
However, inducing multiple acoustic foci (and consequently, nucleating multiple bubbles), which would promote bubble-pair jets is harder to achieve in smaller droplets, as they present less available volume to nucleate multiple bubbles, and require larger input frequencies or highly distorted nonlinear waves to obtain higher wavenumbers and therefore, amplification from higher modes within the droplet.
While employing higher frequencies or distorted waves such as shock waves in applications can be beneficial, nonlinearities are highly attenuated as they propagate through the human body.
Therefore, careful selection of materials, particularly the droplet core, and optimization of the input acoustic driving is needed to obtain piercing jets in sub-micrometric ADV droplets.

\section{Conclusion}
\label{sec:conc}

In this study, we observe ADV with high spatiotemporal resolution and visualize jets from the vapor bubbles nucleated inside the droplet. 
The results obtained here offer new insights into the vapor bubble dynamics in ADV and generally present a broader understanding of ultrasound-induced jetting of cavitation bubbles at the micrometric scale.
Combining the transmission and focusing of the acoustic wave inside the droplet with the radial vapor bubble dynamics, we show that the rapidly varying pressure fields within the droplet are responsible for the formation of high-speed, acoustically-driven jets.
Acoustic focusing leads to the presence of multiple nuclei within the droplet, which also allows us to observe micrometric bubble-pair jets. 
We reason that due to the large variations in the applied acoustic pressure, the growing, evaporating interface of the vapor bubble within the droplet is susceptible to instability, the onset of which may determine the occurrence of a microjet upon bubble collapse.
While difficult to control, harnessing the piercing ability of these jets may be used to enhance membrane permeabilization in biomedical applications, such as tumor ablation, and their creation could be further enhanced by choosing suitable acoustic parameters to create multiple focal spots within the droplet.


\section*{Methods}
\label{sec:Meth}

\begin{figure}[!t]
\centering
\includegraphics[width=0.5\linewidth]{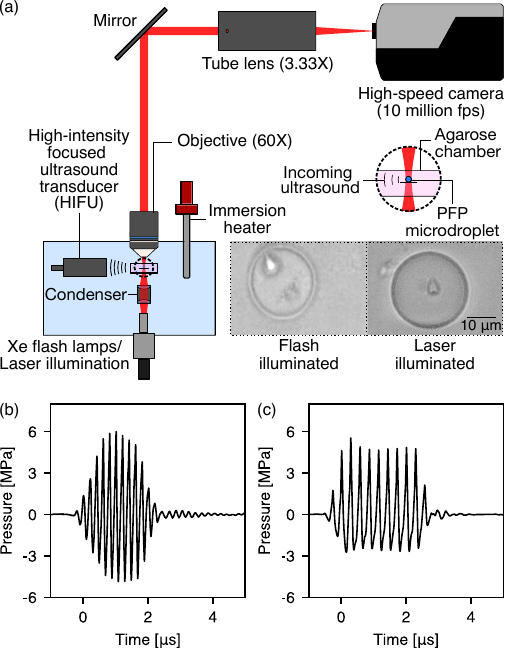}
\caption{\label{fig_8} \textbf{Experimental characterization of ADV.} (a) Schematic of the ultra-high-speed microscopic imaging setup. The zoom-in depicts the PFP droplet placed on the agarose chamber. The inset shows example images of jetting bubbles within droplets illuminated using either the Xe flash lamps \emph{(left)} or the laser \emph{(right)}. (b) Example of an experimental pressure signal measured using a 75-\si{\micro\meter} needle hydrophone for the 5~\si{\mega\hertz} HIFU. (c) Experimental pressure signal for the 3.5~\si{\mega\hertz} HIFU.}
\end{figure}

\subsection*{Experimental setup and image processing}
Perfluoropentane (PFP, \ce{C_5F_12}, ABCR Swiss AG) droplets of radii 1--30~\si{\micro\meter} are prepared by emulsifying 100~\si{\micro\liter} of PFP with a 1\% fluorosurfactant (Capstone\textsuperscript\textregistered~FS-30, Fisher Scientific) solution in a commercial dental mixer (Capsule Mixer, MHC Technology).
The droplets are placed on a 2\% w/v agarose chamber and submerged in a water tank to perform ultra-high-speed microscopy [see Fig.~\ref{fig_8}(a)].
A high-speed camera (HPV-X2, Shimadzu) is coupled to a custom microscope with a 60$\times$ objective (LumPlanFL 60X, Olympus) connected to a 600-\si{\milli\meter} focal length tube lens (TTL-600A, Thorlabs), providing an overall magnification of 200$\times$.
The sample is illuminated, either with a combination of a halogenated light source for live imaging (OSL2, Thorlabs) and Xenon flash lamps (MVS-7010, EG\&G) for video recording, or with a 645-\si{\nano\meter} laser (CAVILUX Smart UHS, Cavitar) that can be used for both. 
The laser illumination was incorporated at a later time in the setup and is preferred to provide a lower exposure time ($t_\mathrm{ex}=20$~\si{\nano\second}) while recording, reducing motion blur within the video.
The droplets are sonicated by two different high-intensity focused ultrasound (HIFU) transducers, one with a center frequency of 3.5~\si{\mega\hertz} (C381-SU, Olympus Panametrics) and the other 5~\si{\mega\hertz} (TXH-5-15, Precision Acoustics). 
The selected transducer is coupled to an amplifier (1020L, E\&I) receiving a single, 10-cycle sine wave pulse of varying amplitude as input from an arbitrary wave generator (LWB420, Teledyne LeCroy).
The experimental pressure wave is measured using a 75-\si{\micro\meter} needle hydrophone (NH0075, Precision Acoustics). 
Examples of such measurements are depicted in Fig.~\ref{fig_8}(b) and (c), and their steady-state cycles are used as inputs for the theoretical modeling of the acoustic wave transmission inside the droplet.
The temperature of the water tank is varied between 36--42\si{\celsius} by a controlled immersion heater.
The relative timings of the camera, the transducer, and the light source are adjusted by a standard delay generator (DG645, Stanford Research Systems). 
Droplet and bubble radii, as well as the progression of the jet tip to evaluate its speed, are measured using ImageJ~\cite{Schneider2012NIHAnalysis}.
Edge detection with Otsu's thresholding is employed where possible~\cite{Otsu1979AHistograms}, while in some cases, the background noise in the image combined with the presence of several interfaces means that manual measurement of these values is required.


\section*{Acknowledgments}
Funding for this work was provided by the Swiss National Science Foundation (Project no. 200567).

\section*{Competing interests}
The authors declare no competing interests.

\section*{Author contributions}

A.P.: Conceptualization, data curation, formal analysis, investigation, methodology, software, visualization, writing - original draft preparation, writing - review \& editing; 
S.F.: Data curation, investigation, methodology, software, writing - review \& editing; 
G.S.: Investigation, methodology, writing - review \& editing; 
O.S.: Conceptualization, funding acquisition, formal analysis, resources, supervision, writing - review \& editing

\section*{Data availability}

The data that supports the findings of this study are available from the corresponding authors upon reasonable request.

\section*{Code availability}

The codes used to simulate the steady-state acoustic wave equation are available as part of the work of Fiorini~\emph{et al.}~\cite{Fiorini2025PositiveVaporization}.
The codes used to evaluate the radial bubble dynamics are available here: \url{https://gitlab.ethz.ch/mfd-group/acoustic-vapor-bubble-dynamics}.

\bibliography{bibliography}

\begin{thebibliography}{10}
\expandafter\ifx\csname url\endcsname\relax
  \def\url#1{\burl{#1}}\fi
\expandafter\ifx\csname urlprefix\endcsname\relax\def\urlprefix{URL }\fi
\providecommand{\bibinfo}[2]{#2}
\providecommand{\eprint}[2][]{\url{#2}}
\providecommand{\doi}[1]{\url{https://doi.org/#1}}
\bibcommenthead

\bibitem{Sehgal1995SonographicMedia.}
\bibinfo{author}{Sehgal, C.~M.}, \bibinfo{author}{Arger, P.~H.} \& \bibinfo{author}{Pugh, C.~R.}
\newblock \bibinfo{title}{{Sonographic enhancement of renal cortex by contrast media.}}
\newblock \emph{\bibinfo{journal}{Journal of Ultrasound in Medicine}} \textbf{\bibinfo{volume}{14}}, \bibinfo{pages}{741--748} (\bibinfo{year}{1995}).

\bibitem{Albrecht1996RenalEchoGen}
\bibinfo{author}{Albrecht, T.} \emph{et~al.}
\newblock \bibinfo{title}{{Renal, Hepatic, and Cardiac Enhancement on Doppler and Gray-Scale Sonograms Obtained with EchoGen}}.
\newblock \emph{\bibinfo{journal}{Academic Radiology}} \textbf{\bibinfo{volume}{3}}, \bibinfo{pages}{198--200} (\bibinfo{year}{1996}).

\bibitem{Forsberg1995ParenchymalAgent.}
\bibinfo{author}{Forsberg, F.}, \bibinfo{author}{Liu, J.~B.}, \bibinfo{author}{Merton, D.~A.}, \bibinfo{author}{Rawool, N.~M.} \& \bibinfo{author}{Goldberg, B.~B.}
\newblock \bibinfo{title}{{Parenchymal enhancement and tumor visualization using a new sonographic contrast agent.}}
\newblock \emph{\bibinfo{journal}{Journal of Ultrasound in Medicine}} \textbf{\bibinfo{volume}{14}}, \bibinfo{pages}{949--957} (\bibinfo{year}{1995}).

\bibitem{kripfgans2000}
\bibinfo{author}{Kripfgans, O.~D.}, \bibinfo{author}{Fowlkes, J.~B.}, \bibinfo{author}{Miller, D.~L.}, \bibinfo{author}{Eldevik, O.~P.} \& \bibinfo{author}{Carson, P.~L.}
\newblock \bibinfo{title}{{Acoustic droplet vaporization for therapeutic and diagnostic applications}}.
\newblock \emph{\bibinfo{journal}{Ultrasound in Med. {\&} Biol.}} \textbf{\bibinfo{volume}{26}}, \bibinfo{pages}{1177--1189} (\bibinfo{year}{2000}).

\bibitem{Harmon2019MinimallyCarcinoma}
\bibinfo{author}{Harmon, J.~S.} \emph{et~al.}
\newblock \bibinfo{title}{{Minimally invasive gas embolization using acoustic droplet vaporization in a rodent model of hepatocellular carcinoma}}.
\newblock \emph{\bibinfo{journal}{Scientific Reports}} \textbf{\bibinfo{volume}{9}} (\bibinfo{year}{2019}).

\bibitem{chen2013}
\bibinfo{author}{Chen, C.~C.} \emph{et~al.}
\newblock \bibinfo{title}{{Targeted drug delivery with focused ultrasound-induced blood-brain barrier opening using acoustically-activated nanodroplets}}.
\newblock \emph{\bibinfo{journal}{Journal of Controlled Release}} \textbf{\bibinfo{volume}{172}}, \bibinfo{pages}{795--804} (\bibinfo{year}{2013}).

\bibitem{sheng2021}
\bibinfo{author}{Sheng, D.}, \bibinfo{author}{Deng, L.}, \bibinfo{author}{Li, P.}, \bibinfo{author}{Wang, Z.} \& \bibinfo{author}{Zhang, Q.}
\newblock \bibinfo{title}{{Perfluorocarbon Nanodroplets with Deep Tumor Penetration and Controlled Drug Delivery for Ultrasound/Fluorescence Imaging Guided Breast Cancer Therapy}}.
\newblock \emph{\bibinfo{journal}{ACS Biomaterials Science and Engineering}} \textbf{\bibinfo{volume}{7}}, \bibinfo{pages}{605--616} (\bibinfo{year}{2021}).

\bibitem{Zhang2011AcousticUltrasound}
\bibinfo{author}{Zhang, M.} \emph{et~al.}
\newblock \bibinfo{title}{{Acoustic Droplet Vaporization for Enhancement of Thermal Ablation by High Intensity Focused Ultrasound}}.
\newblock \emph{\bibinfo{journal}{Academic Radiology}} \textbf{\bibinfo{volume}{18}}, \bibinfo{pages}{1123--1132} (\bibinfo{year}{2011}).

\bibitem{vlai2015}
\bibinfo{author}{Vlaisavljevich, E.} \emph{et~al.}
\newblock \bibinfo{title}{{Effects of Ultrasound Frequency on Nanodroplet-Mediated Histotripsy}}.
\newblock \emph{\bibinfo{journal}{Ultrasound in Medicine and Biology}} \textbf{\bibinfo{volume}{41}}, \bibinfo{pages}{2135--2147} (\bibinfo{year}{2015}).

\bibitem{borden2020}
\bibinfo{author}{Borden, M.~A.}, \bibinfo{author}{Shakya, G.}, \bibinfo{author}{Upadhyay, A.} \& \bibinfo{author}{Song, K.~H.}
\newblock \bibinfo{title}{{Acoustic nanodrops for biomedical applications}}.
\newblock \emph{\bibinfo{journal}{Current Opinion in Colloid and Interface Science}} \textbf{\bibinfo{volume}{50}} (\bibinfo{year}{2020}).

\bibitem{Shakya2024Ultrasound-responsiveDelivery}
\bibinfo{author}{Shakya, G.} \emph{et~al.}
\newblock \bibinfo{title}{{Ultrasound-responsive microbubbles and nanodroplets: A pathway to targeted drug delivery}}.
\newblock \emph{\bibinfo{journal}{Advanced Drug Delivery Reviews}} \textbf{\bibinfo{volume}{206}}, \bibinfo{pages}{115178} (\bibinfo{year}{2024}).

\bibitem{Maeda2015TowardHeterogeneity}
\bibinfo{author}{Maeda, H.}
\newblock \bibinfo{title}{{Toward a full understanding of the EPR effect in primary and metastatic tumors as well as issues related to its heterogeneity}}.
\newblock \emph{\bibinfo{journal}{Advanced Drug Delivery Reviews}} \textbf{\bibinfo{volume}{91}}, \bibinfo{pages}{3--6} (\bibinfo{year}{2015}).

\bibitem{fan2018}
\bibinfo{author}{Fan, C.~H.}, \bibinfo{author}{Lin, Y.~T.}, \bibinfo{author}{Ho, Y.~J.} \& \bibinfo{author}{Yeh, C.~K.}
\newblock \bibinfo{title}{{Spatial-temporal cellular bioeffects from acoustic droplet vaporization}}.
\newblock \emph{\bibinfo{journal}{Theranostics}} \textbf{\bibinfo{volume}{8}}, \bibinfo{pages}{5731--5743} (\bibinfo{year}{2018}).

\bibitem{shpak2014}
\bibinfo{author}{Shpak, O.} \emph{et~al.}
\newblock \bibinfo{title}{{Acoustic droplet vaporization is initiated by superharmonic focusing}}.
\newblock \emph{\bibinfo{journal}{Proceedings of the National Academy of Sciences of the United States of America}} \textbf{\bibinfo{volume}{111}} (\bibinfo{year}{2014}).

\bibitem{lajoinie2021}
\bibinfo{author}{Lajoinie, G.}, \bibinfo{author}{Segers, T.} \& \bibinfo{author}{Versluis, M.}
\newblock \bibinfo{title}{{High-Frequency Acoustic Droplet Vaporization is Initiated by Resonance}}.
\newblock \emph{\bibinfo{journal}{Physical Review Letters}} \textbf{\bibinfo{volume}{126}} (\bibinfo{year}{2021}).

\bibitem{Fiorini2025PositiveVaporization}
\bibinfo{author}{Fiorini, S.}, \bibinfo{author}{Prasanna, A.}, \bibinfo{author}{Shakya, G.}, \bibinfo{author}{Cattaneo, M.} \& \bibinfo{author}{Supponen, O.}
\newblock \bibinfo{title}{{Positive pressure matters in acoustic droplet vaporization}}.
\newblock \emph{\bibinfo{journal}{Physical Review Research}} \textbf{\bibinfo{volume}{7}}, \bibinfo{pages}{023322} (\bibinfo{year}{2025}).

\bibitem{shpak2013c}
\bibinfo{author}{Shpak, O.} \emph{et~al.}
\newblock \bibinfo{title}{{Ultrafast dynamics of the acoustic vaporization of phase-change microdroplets}}.
\newblock \emph{\bibinfo{journal}{The Journal of the Acoustical Society of America}} \textbf{\bibinfo{volume}{134}}, \bibinfo{pages}{1610--1621} (\bibinfo{year}{2013}).

\bibitem{Abeid2024Ultra-high-speedPoint}
\bibinfo{author}{Abeid, B.~A.}, \bibinfo{author}{Fabiilli, M.~L.}, \bibinfo{author}{Estrada, J.~B.} \& \bibinfo{author}{Aliabouzar, M.}
\newblock \bibinfo{title}{{Ultra-high-speed dynamics of acoustic droplet vaporization in soft biomaterials: Effects of viscoelasticity, frequency, and bulk boiling point}}.
\newblock \emph{\bibinfo{journal}{Ultrasonics Sonochemistry}} \textbf{\bibinfo{volume}{103}} (\bibinfo{year}{2024}).

\bibitem{shpak2013}
\bibinfo{author}{Shpak, O.}, \bibinfo{author}{Stricker, L.}, \bibinfo{author}{Versluis, M.} \& \bibinfo{author}{Lohse, D.}
\newblock \bibinfo{title}{{The role of gas in ultrasonically driven vapor bubble growth}}.
\newblock \emph{\bibinfo{journal}{Physics in Medicine and Biology}} \textbf{\bibinfo{volume}{58}}, \bibinfo{pages}{2523--2535} (\bibinfo{year}{2013}).

\bibitem{Ghasemi2022AnShell}
\bibinfo{author}{Ghasemi, M.}, \bibinfo{author}{Yu, A.~C.} \& \bibinfo{author}{Sivaloganathan, S.}
\newblock \bibinfo{title}{{An enhanced, rational model to study acoustic vaporization dynamics of a bubble encapsulated within a nonlinearly elastic shell}}.
\newblock \emph{\bibinfo{journal}{Ultrasonics Sonochemistry}} \textbf{\bibinfo{volume}{83}} (\bibinfo{year}{2022}).

\bibitem{Doinikov2014VaporizationValidation}
\bibinfo{author}{Doinikov, A.~A.}, \bibinfo{author}{Sheeran, P.~S.}, \bibinfo{author}{Bouakaz, A.} \& \bibinfo{author}{Dayton, P.~A.}
\newblock \bibinfo{title}{{Vaporization dynamics of volatile perfluorocarbon droplets: A theoretical model and in vitro validation}}.
\newblock \emph{\bibinfo{journal}{Medical Physics}} \textbf{\bibinfo{volume}{41}} (\bibinfo{year}{2014}).

\bibitem{Sankin2005ShockBubbles}
\bibinfo{author}{Sankin, G.~N.}, \bibinfo{author}{Simmons, W.~N.}, \bibinfo{author}{Zhu, S.~L.} \& \bibinfo{author}{Zhong, P.}
\newblock \bibinfo{title}{{Shock wave interaction with laser-generated single bubbles}}.
\newblock \emph{\bibinfo{journal}{Physical Review Letters}} \textbf{\bibinfo{volume}{95}} (\bibinfo{year}{2005}).

\bibitem{Brujan2017JetsBoundary}
\bibinfo{author}{Brujan, E.~A.}
\newblock \bibinfo{title}{{Jets from pulsed-ultrasound-induced cavitation bubbles near a rigid boundary}}.
\newblock \emph{\bibinfo{journal}{Journal of Physics D: Applied Physics}} \textbf{\bibinfo{volume}{50}} (\bibinfo{year}{2017}).

\bibitem{blake1981}
\bibinfo{author}{Blake, J.~R.} \& \bibinfo{author}{Cerone, P.}
\newblock \bibinfo{title}{{A Note on the Impulse Due to a Vapour Bubble Near a Boundary}}.
\newblock \emph{\bibinfo{journal}{J. Austral. Math. Soc. (Series B)}} \textbf{\bibinfo{volume}{23}}, \bibinfo{pages}{383--393} (\bibinfo{year}{1982}).

\bibitem{tomita2002}
\bibinfo{author}{Tomita, Y.}, \bibinfo{author}{Robinson, P.~B.}, \bibinfo{author}{Tong, R.~P.} \& \bibinfo{author}{Blake, J.~R.}
\newblock \bibinfo{title}{{Growth and collapse of cavitation bubbles near a curved rigid boundary}}.
\newblock \emph{\bibinfo{journal}{Journal of Fluid Mechanics}} \textbf{\bibinfo{volume}{466}}, \bibinfo{pages}{259--283} (\bibinfo{year}{2002}).

\bibitem{Blake1981GrowthSurface}
\bibinfo{author}{Blake, J.~R.} \& \bibinfo{author}{Gibson, D.~C.}
\newblock \bibinfo{title}{{Growth and Collapse of A Vapour Cavity Near A Free Surface}}.
\newblock \emph{\bibinfo{journal}{Journal of Fluid Mechanics}} \textbf{\bibinfo{volume}{111}}, \bibinfo{pages}{123--140} (\bibinfo{year}{1981}).

\bibitem{Supponen2016ScalingBubbles}
\bibinfo{author}{Supponen, O.} \emph{et~al.}
\newblock \bibinfo{title}{{Scaling laws for jets of single cavitation bubbles}}.
\newblock \emph{\bibinfo{journal}{Journal of Fluid Mechanics}} \textbf{\bibinfo{volume}{802}}, \bibinfo{pages}{263--293} (\bibinfo{year}{2016}).

\bibitem{han2015}
\bibinfo{author}{Han, B.} \emph{et~al.}
\newblock \bibinfo{title}{{Dynamics of laser-induced bubble pairs}}.
\newblock \emph{\bibinfo{journal}{Journal of Fluid Mechanics}} \textbf{\bibinfo{volume}{771}}, \bibinfo{pages}{706--742} (\bibinfo{year}{2015}).

\bibitem{Mishra2022Flow-focusingBubbles}
\bibinfo{author}{Mishra, A.} \emph{et~al.}
\newblock \bibinfo{title}{{Flow-focusing from interacting cavitation bubbles}}.
\newblock \emph{\bibinfo{journal}{Physical Review Fluids}} \textbf{\bibinfo{volume}{7}}, \bibinfo{pages}{110502} (\bibinfo{year}{2022}).

\bibitem{Blake2015CavitationApplications}
\bibinfo{author}{Blake, J.~R.}, \bibinfo{author}{Leppinen, D.~M.} \& \bibinfo{author}{Wang, Q.}
\newblock \bibinfo{title}{{Cavitation and bubble dynamics: The Kelvin impulse and its applications}}.
\newblock \emph{\bibinfo{journal}{Interface Focus}} \textbf{\bibinfo{volume}{5}}, \bibinfo{pages}{1--15} (\bibinfo{year}{2015}).

\bibitem{Brujan2005JetUltrasound}
\bibinfo{author}{Brujan, E.~A.}, \bibinfo{author}{Ikeda, T.} \& \bibinfo{author}{Matsumoto, Y.}
\newblock \bibinfo{title}{{Jet formation and shock wave emission during collapse of ultrasound-induced cavitation bubbles and their role in the therapeutic applications of high-intensity focused ultrasound}}.
\newblock \emph{\bibinfo{journal}{Physics in Medicine and Biology}} \textbf{\bibinfo{volume}{50}}, \bibinfo{pages}{4797--4809} (\bibinfo{year}{2005}).

\bibitem{gerold2012}
\bibinfo{author}{Gerold, B.} \emph{et~al.}
\newblock \bibinfo{title}{{Directed jetting from collapsing cavities exposed to focused ultrasound}}.
\newblock \emph{\bibinfo{journal}{Applied Physics Letters}} \textbf{\bibinfo{volume}{100}} (\bibinfo{year}{2012}).

\bibitem{rossello2018}
\bibinfo{author}{Rossell{\'{o}}, J.~M.} \emph{et~al.}
\newblock \bibinfo{title}{{Acoustically induced bubble jets}}.
\newblock \emph{\bibinfo{journal}{Physics of Fluids}} \textbf{\bibinfo{volume}{30}} (\bibinfo{year}{2018}).

\bibitem{Banine2011PhysicalMicrolithography}
\bibinfo{author}{Banine, V.~Y.}, \bibinfo{author}{Koshelev, K.~N.} \& \bibinfo{author}{Swinkels, G.~H.}
\newblock \bibinfo{title}{{Physical processes in EUV sources for microlithography}}.
\newblock \emph{\bibinfo{journal}{Journal of Physics D: Applied Physics}} \textbf{\bibinfo{volume}{44}} (\bibinfo{year}{2011}).

\bibitem{Lee2022AFiber}
\bibinfo{author}{Lee, H.} \emph{et~al.}
\newblock \bibinfo{title}{{A laser-driven optical atomizer: photothermal generation and transport of zeptoliter-droplets along a carbon nanotube deposited hollow optical fiber}}.
\newblock \emph{\bibinfo{journal}{Nanoscale}} \textbf{\bibinfo{volume}{14}}, \bibinfo{pages}{5138--5146} (\bibinfo{year}{2022}).

\bibitem{Mei2015AtmosphericSystem}
\bibinfo{author}{Mei, L.} \& \bibinfo{author}{Brydegaard, M.}
\newblock \bibinfo{title}{{Atmospheric aerosol monitoring by an elastic Scheimpflug lidar system}}.
\newblock \emph{\bibinfo{journal}{Optics Express}} \textbf{\bibinfo{volume}{23}}, \bibinfo{pages}{A1613} (\bibinfo{year}{2015}).

\bibitem{obreschkow2006}
\bibinfo{author}{Obreschkow, D.} \emph{et~al.}
\newblock \bibinfo{title}{{Cavitation bubble dynamics inside liquid drops in microgravity}}.
\newblock \emph{\bibinfo{journal}{Physical Review Letters}} \textbf{\bibinfo{volume}{97}} (\bibinfo{year}{2006}).

\bibitem{Rossello2023BubbleDroplet}
\bibinfo{author}{Rossell{\'{o}}, J.~M.}, \bibinfo{author}{Reese, H.}, \bibinfo{author}{Raman, K.~A.} \& \bibinfo{author}{Ohl, C.~D.}
\newblock \bibinfo{title}{{Bubble nucleation and jetting inside a millimetric droplet}}.
\newblock \emph{\bibinfo{journal}{Journal of Fluid Mechanics}} \textbf{\bibinfo{volume}{968}} (\bibinfo{year}{2023}).

\bibitem{Li2024CavitationFluid}
\bibinfo{author}{Li, S.}, \bibinfo{author}{Zhao, Z.}, \bibinfo{author}{Zhang, A.~M.} \& \bibinfo{author}{Han, R.}
\newblock \bibinfo{title}{{Cavitation bubble dynamics inside a droplet suspended in a different host fluid}}.
\newblock \emph{\bibinfo{journal}{Journal of Fluid Mechanics}} \textbf{\bibinfo{volume}{979}} (\bibinfo{year}{2024}).

\bibitem{sankin2010}
\bibinfo{author}{Sankin, G.~N.}, \bibinfo{author}{Yuan, F.} \& \bibinfo{author}{Zhong, P.}
\newblock \bibinfo{title}{{Pulsating tandem microbubble for localized and directional single-cell membrane poration}}.
\newblock \emph{\bibinfo{journal}{Physical Review Letters}} \textbf{\bibinfo{volume}{105}} (\bibinfo{year}{2010}).

\bibitem{Tomita2017PulsedTimes}
\bibinfo{author}{Tomita, Y.} \& \bibinfo{author}{Sato, K.}
\newblock \bibinfo{title}{{Pulsed jets driven by two interacting cavitation bubbles produced at different times}}.
\newblock \emph{\bibinfo{journal}{Journal of Fluid Mechanics}} \textbf{\bibinfo{volume}{819}}, \bibinfo{pages}{465--493} (\bibinfo{year}{2017}).

\bibitem{Ohl2006SonoporationBubbles}
\bibinfo{author}{Ohl, C.~D.} \emph{et~al.}
\newblock \bibinfo{title}{{Sonoporation from jetting cavitation bubbles}}.
\newblock \emph{\bibinfo{journal}{Biophysical Journal}} \textbf{\bibinfo{volume}{91}}, \bibinfo{pages}{4285--4295} (\bibinfo{year}{2006}).

\bibitem{Qin2021PredictingVaporization}
\bibinfo{author}{Qin, D.}, \bibinfo{author}{Zou, Q.}, \bibinfo{author}{Lei, S.}, \bibinfo{author}{Wang, W.} \& \bibinfo{author}{Li, Z.}
\newblock \bibinfo{title}{{Predicting initial nucleation events occurred in a metastable nanodroplet during acoustic droplet vaporization}}.
\newblock \emph{\bibinfo{journal}{Ultrasonics Sonochemistry}} \textbf{\bibinfo{volume}{75}} (\bibinfo{year}{2021}).

\bibitem{hao1999}
\bibinfo{author}{Hao, Y.} \& \bibinfo{author}{Prosperetti, A.}
\newblock \bibinfo{title}{{The dynamics of vapor bubbles in acoustic pressure fields}}.
\newblock \emph{\bibinfo{journal}{Physics of Fluids}} \textbf{\bibinfo{volume}{11}}, \bibinfo{pages}{2008--2019} (\bibinfo{year}{1999}).

\bibitem{stricker2011}
\bibinfo{author}{Stricker, L.}, \bibinfo{author}{Prosperetti, A.} \& \bibinfo{author}{Lohse, D.}
\newblock \bibinfo{title}{{Validation of an approximate model for the thermal behavior in acoustically driven bubbles}}.
\newblock \emph{\bibinfo{journal}{The Journal of the Acoustical Society of America}} \textbf{\bibinfo{volume}{130}}, \bibinfo{pages}{3243--3251} (\bibinfo{year}{2011}).

\bibitem{Prosperetti1978Vapour-bubbleLiquid}
\bibinfo{author}{Prosperetti, A.} \& \bibinfo{author}{Plesset, M.~S.}
\newblock \bibinfo{title}{{Vapour-bubble growth in a superheated liquid}}.
\newblock \emph{\bibinfo{journal}{Journal of Fluid Mechanics}} \textbf{\bibinfo{volume}{85}}, \bibinfo{pages}{349--368} (\bibinfo{year}{1978}).

\bibitem{longuet1995}
\bibinfo{author}{Longuet-Higgins, M.~S.} \& \bibinfo{author}{Oguz, H.}
\newblock \bibinfo{title}{{Critical microjets in collapsing cavities}}.
\newblock \emph{\bibinfo{journal}{Journal of Fluid Mechanics}} \textbf{\bibinfo{volume}{290}}, \bibinfo{pages}{183--201} (\bibinfo{year}{1995}).

\bibitem{Reese2024CavitationSurfaces}
\bibinfo{author}{Reese, H.}, \bibinfo{author}{Ohl, C.~D.} \& \bibinfo{author}{Rossell{\'{o}}, J.~M.}
\newblock \bibinfo{title}{{Cavitation and jetting from shock wave refocusing near convex liquid surfaces}}.
\newblock \emph{\bibinfo{journal}{International Journal of Multiphase Flow}} \textbf{\bibinfo{volume}{175}} (\bibinfo{year}{2024}).

\bibitem{Chew2011InteractionField}
\bibinfo{author}{Chew, L.~W.}, \bibinfo{author}{Klaseboer, E.}, \bibinfo{author}{Ohl, S.~W.} \& \bibinfo{author}{Khoo, B.~C.}
\newblock \bibinfo{title}{{Interaction of two differently sized oscillating bubbles in a free field}}.
\newblock \emph{\bibinfo{journal}{Physical Review E - Statistical, Nonlinear, and Soft Matter Physics}} \textbf{\bibinfo{volume}{84}} (\bibinfo{year}{2011}).

\bibitem{wong2011}
\bibinfo{author}{Wong, Z.~Z.}, \bibinfo{author}{Kripfgans, O.~D.}, \bibinfo{author}{Qamar, A.}, \bibinfo{author}{Fowlkes, J.~B.} \& \bibinfo{author}{Bull, J.~L.}
\newblock \bibinfo{title}{{Bubble evolution in acoustic droplet vaporization at physiological temperature via ultra-high speed imaging}}.
\newblock \emph{\bibinfo{journal}{Soft Matter}} \textbf{\bibinfo{volume}{7}}, \bibinfo{pages}{4009--4016} (\bibinfo{year}{2011}).

\bibitem{Shepherd1982RapidLimit}
\bibinfo{author}{Shepherd, J.~E.} \& \bibinfo{author}{Sturtevant, B.}
\newblock \bibinfo{title}{{Rapid evaporation at the superheat limit}}.
\newblock \emph{\bibinfo{journal}{Journal of Fluid Mechanics}} \textbf{\bibinfo{volume}{121}}, \bibinfo{pages}{379--402} (\bibinfo{year}{1982}).

\bibitem{Frost1986EffectsLimit}
\bibinfo{author}{Frost, D.} \& \bibinfo{author}{Sturtevant, B.}
\newblock \bibinfo{title}{{Effects of Ambient Pressure on the Instability of a Liquid Boiling Explosively at the Superheat Limit}}.
\newblock \emph{\bibinfo{journal}{Journal of Heat Transfer}} \textbf{\bibinfo{volume}{108}}, \bibinfo{pages}{418--424} (\bibinfo{year}{1986}).

\bibitem{Xie1991EvaporativeDroplets}
\bibinfo{author}{Xie, J.-G.}, \bibinfo{author}{Ruekgauer, T.~E.}, \bibinfo{author}{Armstrong, R.~L.} \& \bibinfo{author}{Pinnick, R.~G.}
\newblock \bibinfo{title}{{Evaporative Instability in Pulsed Laser-Heated Droplets}}.
\newblock \emph{\bibinfo{journal}{Physical Review Letters}} \textbf{\bibinfo{volume}{66}}, \bibinfo{pages}{2988--2991} (\bibinfo{year}{1991}).

\bibitem{Apfel1975AcousticallyDroplets}
\bibinfo{author}{Apfel, R.~E.} \& \bibinfo{author}{Harbison, J.~P.}
\newblock \bibinfo{title}{{Acoustically induced explosions of superheated droplets}}.
\newblock \emph{\bibinfo{journal}{The Journal of the Acoustical Society of America}} \textbf{\bibinfo{volume}{57}}, \bibinfo{pages}{1371--1373} (\bibinfo{year}{1975}).

\bibitem{Frost1989ExperimentsWave}
\bibinfo{author}{Frost, D.~L.}
\newblock \bibinfo{title}{{Experiments in Fluids Initiation of explosive boiling of a droplet with a shock wave}}.
\newblock \emph{\bibinfo{journal}{Experiments in Fluids}} \textbf{\bibinfo{volume}{8}}, \bibinfo{pages}{121--128} (\bibinfo{year}{1989}).

\bibitem{Palmer1976ThePressure}
\bibinfo{author}{Palmer, H.~J.}
\newblock \bibinfo{title}{{The hydrodynamic stability of rapidly evaporating liquids at reduced pressure}}.
\newblock \emph{\bibinfo{journal}{Journal of Fluid Mechanics}} \textbf{\bibinfo{volume}{75}}, \bibinfo{pages}{487--511} (\bibinfo{year}{1976}).

\bibitem{aliabouzar2018}
\bibinfo{author}{Aliabouzar, M.}, \bibinfo{author}{Kumar, K.~N.} \& \bibinfo{author}{Sarkar, K.}
\newblock \bibinfo{title}{{Acoustic vaporization threshold of lipid-coated perfluoropentane droplets}}.
\newblock \emph{\bibinfo{journal}{The Journal of the Acoustical Society of America}} \textbf{\bibinfo{volume}{143}}, \bibinfo{pages}{2001--2012} (\bibinfo{year}{2018}).

\bibitem{Schneider2012NIHAnalysis}
\bibinfo{author}{Schneider, C.~A.}, \bibinfo{author}{Rasband, W.~S.} \& \bibinfo{author}{Eliceiri, K.~W.}
\newblock \bibinfo{title}{{NIH Image to ImageJ: 25 years of image analysis}}.
\newblock \emph{\bibinfo{journal}{Nature Methods}} \textbf{\bibinfo{volume}{9}}, \bibinfo{pages}{671--675} (\bibinfo{year}{2012}).

\bibitem{Otsu1979AHistograms}
\bibinfo{author}{Otsu, N.}
\newblock \bibinfo{title}{{A threshold selection method from gray-level histograms}}.
\newblock \emph{\bibinfo{journal}{IEEE Transactions on Systems, Manm and Cybernetics}} \textbf{\bibinfo{volume}{SMC. 9}}, \bibinfo{pages}{62--66} (\bibinfo{year}{1979}).

\end{thebibliography}

\end{document}